\documentclass{WileyMSP-template}
\usepackage{subfigure}
\usepackage{multirow}
\usepackage{amsmath}

\usepackage{makecell}
\usepackage{textcomp}
\usepackage{gensymb}

\usepackage{cite}
\begin{document}
	
	\pdfminorversion=7
	\pagestyle{fancy}
	\rhead{\includegraphics[width=2.5cm]{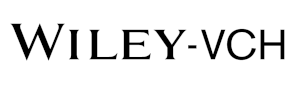}}

	\title{Computing-heightened low-cost high-dimensional controlled-SUM gates}
	
	\maketitle
	
	
\author{Zhi-Guo Fan$^{1*}$}
\author{Zhuo-Ya Bai$^{2}$}
\author{Qiu-Lin Tan$^{1}$}
\author{Fang-Fang Du$^{1*}$}

	
	
	\begin{affiliations}

	$^{1}$Key Laboratory of Micro/nano Devices and Systems, Ministry of Education, North University of China, Tai Yuan 030051, China\\
	Email Address: zhiguo\_fan2023@163.com; duff1987@126.com

	$^{2}$Beijing National Research Center for Information Science and Technology, Department of Electronic Engineering, Tsinghua University, Beijing 100084, 
	China\\

	\end{affiliations}

	

\begin{abstract}
   Qudit-based quantum gates offer several advantages over qubit-based counterparts,
   such as higher information density, the ability to address more complex problems, and richer quantum operations.
   In this paper, we present three realistic protocols for implementing a 4$\times$4-dimensional (16D) two-qudit controlled-SUM (CSUM) gate, where the 4D control qudit and 4D target qudit  are encoded in the polarization degree of freedom (DoF) and spatial DoF of  two photons, respectively. The first protocol is implemented exclusively using linear optical elements without  auxiliary resources, making it feasible with current optical technologies and achieving an efficiency of 1/9. The second protocol utilizes photon scattering by a microcavity-quantum-dot system, enabling the 16D CSUM gate to operate deterministically without postselection.
   The third protocol introduces an error-heralded mechanism based on the second protocol, theoretically achieving unity fidelity.
   Moreover, all protocols operate without ancillary photons,
   offering the advantages of compact circuits and  low cost while
   further promoting the development of  high-dimensional quantum computation.
\end{abstract}
	
	\section{Introduction}   \label{sec1}
Quantum computing is increasingly recognized for its potential advantages over classical computing \cite{1,2,4,5,6}. Quantum logic gates are fundamental units to quantum computing processes \cite{802,803,310,901,902,801,804,805,806,807}.  High-dimensional (HD) qudits, which extend two-dimensional qubits \cite{8,9,10,11,12,13}, significantly enhance the channel capacity of quantum resources \cite{16,17}, enabling HD quantum information transmission \cite{18,302,20,303,304,311,21} and potentially facilitating fault-tolerant quantum computing \cite{22} and distributed quantum networks \cite{24,25,26,27}. In HD space, qudit systems offer great flexibility for information storage and processing, including simplified quantum gates \cite{28}, improved information security \cite{29,305,306,30,31,32,33}, and the exploration of unique quantum properties \cite{34,35}. 
Beyond quantum computing, qudits
offer benefits in quantum communication due to their superior noise resilience \cite{36,37,38,39,40,41} and their ability to support higher key rates \cite{308,44,45,46,307}.

Reliable quantum computing necessitates high-fidelity quantum logic gates. The controlled-NOT (CNOT) gate, in particular, is extensively utilized in quantum computing for applications, such as quantum algorithms \cite{48,49,51}, error correction \cite{22}, arithmetic operations \cite{53,54}, and fault-tolerant computing \cite{55,56,57}. As circuit complexity grows, the limitations of traditional low-dimensional CNOT gates become more pronounced, increasing the demand for  design of HD counterpart gates, i,e., controlled-SUM (CSUM) gates.
In the $d\times d$-D ($d>2$) Hilbert space, 
the $d\times d$-D  CSUM  gate can be expressed as $U_{\text{CSUM}}^d|c,t\rangle=|c,(c+t)\%d\rangle$, where $c,t\in\{0,1,...,d-1\}$.
It is evident that the CNOT gate is a special case of the CSUM gate when 
$d=2$.
Especially, the $4\times 4$-D (16D) CSUM gate ($d$ = 4) performs the following transformation:
\[
\begin{aligned}
	|0,0\rangle &\rightarrow |0,0\rangle, |0,1\rangle \rightarrow |0,1\rangle, |0,2\rangle \rightarrow |0,2\rangle, |0,3\rangle \rightarrow |0,3\rangle, \\
	|1,0\rangle &\rightarrow |1,1\rangle, |1,1\rangle \rightarrow |1,2\rangle, |1,2\rangle \rightarrow |1,3\rangle, |1,3\rangle \rightarrow |1,0\rangle, \\
	|2,0\rangle &\rightarrow |2,2\rangle, |2,1\rangle \rightarrow |2,3\rangle, |2,2\rangle \rightarrow |2,0\rangle, |2,3\rangle \rightarrow |2,1\rangle, \\
	|3,0\rangle &\rightarrow |3,3\rangle, |3,1\rangle \rightarrow |3,0\rangle, |3,2\rangle \rightarrow |3,1\rangle, |3,3\rangle \rightarrow |3,2\rangle.
\end{aligned}
\]

Photon systems possess inherent immunity to decoherence and offer multiple accessible degrees of freedom (DoF), making them well-suited for encoding qudit information \cite{1000,1001,1002,1003,1004,1005}.
These advantages make the  HD CSUM gate implemented exclusively by linear optical elements easier to realize.
Recently, many CSUM gates have been proposed in quantum information processing (QIP).
In 2019, Imany \emph{et al.} proposed a $3\times3$-D CSUM gate with a fidelity exceeding 0.90, utilizing time and frequency DoF for encoding \cite{60}.
In 2020, Gao \emph{et al.} implemented a $3\times3$-D CSUM gate with the assistance of a three-photon entangled state, achieving an efficiency of 1/152  \cite{59}.
In 2022, Su \emph{et al.}  reported  a $3\times3$-D
hybrid  CSUM gate with one superconducting qutrit and a cat-state qutrit \cite{PhysRevA.105.042434}. 
In 2024, a $2\times4$-D polarization-spatial CSUM gate, with an average fidelity and efficiency exceeding 0.99, was realized by Meng \emph{et al.}\cite{PhysRevA.109.022612}.
Most qudit-based gates offer several advantages over qubit-based counterparts, such as shorter computation times, reduced resource requirements, greater availability, and the capacity to address more complex problems. Overall, research on HD quantum gates broadens the current quantum computing framework \cite{61}.

In this paper, we present three useful protocols to set up the 16D two-qudit CSUM gate, where the 4D control qudit and 4D target qudit of the CSUM gate are encoded in the polarization DoF and spatial DoF of two photons, respectively. The first protocol employs linear optical elements achieving an efficiency of 1/9, which beats the $3\times3$-D CSUM gate with the efficiency  of 1/152 assisted by
the three-photon entangled state \cite{59}. The second protocol is implemented through photon scattering off a microcavity-quantum-dot     (microcavity-QD) system, where it operates deterministically. The third protocol introduces an error-heralded framework on the foundation of the second protocol, elevating the fidelity of the 16D CSUM gate near unity, in principle. Moreover, each protocol can requires no ancillary photons, offering the advantages of compact circuits and  low cost.


\section{The 16D CSUM gate implemented with linear optics}   \label{sec2}

\begin{figure}
	\centering
	\begin{center}
		\centering
		\includegraphics[width=0.65\linewidth]{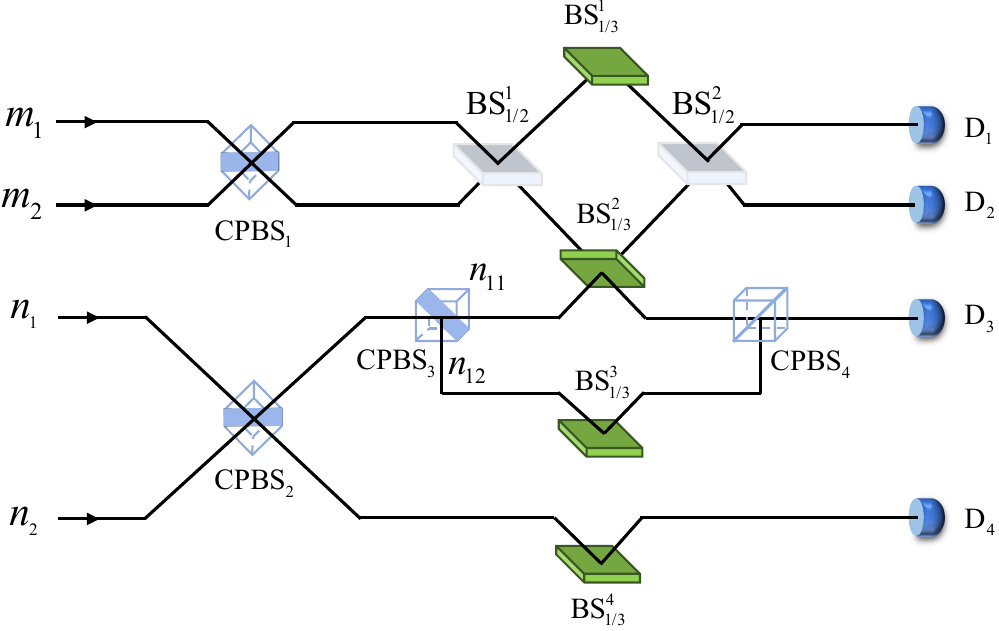}
		\caption{Schematic diagram of  16D CSUM gate  implemented with linear optics.
			CPBS denotes a circular polarization beam splitter, which transmit (reflect) the  $|R\rangle-$ ($|L\rangle$-)  polarization state of the photon.
			BS$_{1/2}$ is a 1/2-reflectivity beam splitter.
			BS$_{1/3}$ is a 1/3-reflectivity beam splitter.
			D is a single-photon detector.}\label{fig1}
	\end{center}
\end{figure}

The schematic diagram of the first protocol for the 16D CSUM gate implemented with linear optics is illustrated in Figure \ref{fig1}.
The initial states of the two photons, labeled $M$ and $N$, in both polarization and spatial DoFs, are
\begin{eqnarray} \label{eq7}
	|\varphi\rangle_{MN}^{P}&=&
	(\alpha_{1}|LL\rangle+\beta_{1}|LR\rangle+\gamma_{1}|RL\rangle+\delta_{1}|RR\rangle)_{MN},\nonumber\\
	|\varphi\rangle_{MN}^{S}&=&
	(\alpha_{2}|m_{1}n_{1}\rangle+\beta_{2}|m_{1}n_{2}\rangle+\gamma_{2}|m_{2}n_{1}\rangle+\delta_{2}|m_{2}n_{2}\rangle)_{MN},
\end{eqnarray}
where the coefficients adhere the normalization condition $|\alpha_{i}|^2 + |\beta_{i}|^2 + |\gamma_{i}|^2 + |\delta_{i}|^2 = 1$ for $i = 1, 2$.  Thus, the initial states of the entire system can be expressed as $|\Psi_0\rangle = |\varphi\rangle_{MN}^{P} \otimes |\varphi\rangle_{MN}^{S}$.

Firstly, two photons $M$ and $N$ in spatial modes $m_{1}$, $m_{2}$, $n_{1}$, and $n_{2}$ pass through respective circular polarized beam splitters (CPBS$_{1}$ and CPBS$_{2}$), which transmit (reflect) the $|R\rangle$ ($|L\rangle$) polarization state of the photon. Owing to the property of the CPBS, only the spatial modes entangled with $|R\rangle$-polarized photons are exchanged, resulting in
\begin{eqnarray} \label{eq8}
	|\Psi_1\rangle &=&
	\alpha_1 |LL\rangle_{MN} \otimes (\alpha_{2} |m_1 n_1\rangle + \beta_2 |m_1 n_2\rangle + \gamma_{2} |m_2 n_1\rangle+ \delta_{2} |m_2 n_2\rangle)\nonumber\\
	&&+ \beta_{1} |LR\rangle_{MN}\otimes (\alpha_{2} |m_1 n_2\rangle+ \beta_2 |m_1 n_1\rangle+ \gamma_{2} |m_2 n_2\rangle + \delta_{2} |m_2 n_1\rangle) \nonumber\\
	&&+ \gamma_{1} |RL\rangle_{MN}\otimes (\alpha_{2} |m_2 n_1\rangle + \beta_2 |m_2 n_2\rangle + \gamma_{2} |m_1 n_1\rangle + \delta_{2} |m_1 n_2\rangle)\nonumber\\
	&&+ \delta_{1} |RR\rangle_{MN} \otimes (\alpha_{2} |m_2 n_2\rangle + \beta_2 |m_2 n_1\rangle + \gamma_{2} |m_1 n_2\rangle + \delta_{2} |m_1 n_1\rangle).
\end{eqnarray}

Secondly,  for the spatial mode of photon $M$,
an interferometric network is constructed using two 1/2-reflectivity beam splitters (BS\(_{1/2}^1\) and BS\(_{1/2}^2\)), while two 1/3-reflectivity beam splitters (BS\(_{1/3}^1\) and BS\(_{1/3}^2\)) are positioned on the two arms of the interferometer, respectively.
The BS$_{1/2}$ applies the operation to two spatial modes of the photon $M$, i.e., $|m_{1}\rangle = (|m_{1}\rangle + |m_{2}\rangle)/\sqrt{2}$ and $|m_{2}\rangle = (|m_{1}\rangle - |m_{2}\rangle)/\sqrt{2}$.
For photon \( N \), after passing through the CPBS$_{3}$, its spatial mode \( n_{1} \) is split into the spatial modes \( n_{11} \) and \( n_{12} \). These modes are subsequently recombined into the spatial mode \(n_{1} \) by CPBS$_{4}$.
The photon \( M \) in the spatial mode \( m_{2} \) and the photon \( N \) in the spatial mode \( n_{11} \) simultaneously pass through the BS$_{1/3}^{2}$. 
This induces interference between the two photons, resulting in a 
\( \pi \)-phase shift in the interferometric arms \cite{PhysRevA.66.024308}.
It is noteworthy that the BS$_{1/3}^{1}$, BS$_{1/3}^{3}$, and BS$_{1/3}^{4}$ in proper order the  $m_{1}$, $n_{12}$, and $n_{2}$ spatial modes do not participate in the interference between two photons but serves to average the photon loss.
After these operation, the state of the entire system is evolved to
\begin{eqnarray} \label{eq9}
	|\Psi_2\rangle&=& \frac{1}{3}\{\alpha_1|LL\rangle_{MN}\otimes(\alpha_2|m_{1}n_1\rangle+\beta_2|m_{1}n_2\rangle+\gamma_2|m_{2}n_1\rangle+\delta_2|m_{2}n_2\rangle)\nonumber\\
	&&+\beta_{1}|LR\rangle_{MN}\otimes(\alpha_2|m_{1}n_2\rangle+\beta_2|m_{2}n_1\rangle+\gamma_2|m_{2}n_2\rangle+\delta_2|m_{1}n_1\rangle)\nonumber\\
	&&+\gamma_{1}|RL\rangle_{MN}\otimes(\alpha_2|m_{2}n_1\rangle+\beta_2|m_{2}n_2\rangle+\gamma_2|m_{1}n_1\rangle+\delta_2|m_{1}n_2\rangle)\nonumber\\
	&&+\delta_{1}|RR\rangle_{MN} \otimes(\alpha_2|m_{2}n_2\rangle+\beta_2||m_{1}n_1\rangle+\gamma_2|m_{1}n_2\rangle+\delta_2|m_{2}n_1\rangle)\}.
\end{eqnarray}
In Equation (\ref{eq9}), we retain only the photon-number-conserving terms detectable by photon detector D, while omitting the failure modes of the CSUM gate where the output ports of the single photon do not each contain exactly one photon.
Comparing Equation (\ref{eq8}) with
Equation (\ref{eq9}), it can be seen that,
when photon $N$  is in the $ |Rn_{1}\rangle_{N} $ state, the $|m_{1}\rangle$ component of photon $M$ is converted into $|m_{2}\rangle$ one
and vice versa.
In all other cases, the states of two photons $N$ and $M$ remain unchanged.

The polarization states of two photons $M$ and $N$ are served as the 4D control qudit, that is, $|LL\rangle \rightarrow |\bar{0}\rangle_{c}, |LR\rangle \rightarrow |\bar{1}\rangle_{c}, |RL\rangle \rightarrow |\bar{2}\rangle_{c}$, and $|RR\rangle \rightarrow |\bar{3}\rangle_{c}$. Simultaneously, the spatial states of two photons $M$ and $N$ are worked as 4D target qudit, that is, $|m_{1}n_{1}\rangle \rightarrow |\bar{0}\rangle_{t}, |m_{1}n_{2}\rangle \rightarrow |\bar{1}\rangle_{t}, |m_{2}n_{1}\rangle \rightarrow |\bar{2}\rangle_{t}, |m_{2}n_{2}\rangle \rightarrow |\bar{3}\rangle_{t}$. Consequently, the  16D CSUM gate can be successfully implemented with linear optical elements, i.e.,
\begin{eqnarray} \label{eq10}
	|\rm CSUM\rangle&=&
	\frac{1}{3}\{\alpha_1|\bar{0}\rangle_{c}\otimes(\alpha_2|\bar{0}\rangle_{t}+\beta_2|\bar{1}\rangle_{t}
	+\gamma_2|\bar{2}\rangle_{t}+\delta_2|\bar{3}\rangle_{t})\nonumber\\
	&&+\beta_{1}|\bar{1}\rangle_{c}\otimes(\alpha_2|\bar{1}\rangle_{t}+\beta_2|\bar{2}\rangle_{t}
	+\gamma_2|\bar{3}\rangle_{t}+\delta_2|\bar{0}\rangle_{t})\nonumber\\
	&&+\gamma_{1}|\bar{2}\rangle_{c}\otimes(\alpha_2|\bar{2}\rangle_{t}+\beta_2|\bar{3}\rangle_{t}
	+\gamma_2|\bar{0}\rangle_{t}+\delta_2|\bar{1}\rangle_{t})\nonumber\\
	&&+\delta_{1}|\bar{3}\rangle_{c} \otimes(\alpha_2|\bar{3}\rangle_{t}+\beta_2|\bar{0}\rangle_{t}
	+\gamma_2|\bar{1}\rangle_{t}+\delta_2|\bar{2}\rangle_{t})\}.
\end{eqnarray}
It is evident that  the efficiency $\eta_1$ of 16D CSUM gate with linear optics is  1/9.
We have completed the first protocol by encoding information in the spatial and polarized DoFs of two flying photons,  resulting in significant resource savings.

\section{The deterministic  16D CSUM gate assisted by two one-sided microcavity-QD systems}  \label{sec3}



\begin{figure*}
	\centering
	\begin{center}
		\centering
		\includegraphics[width=1\linewidth]{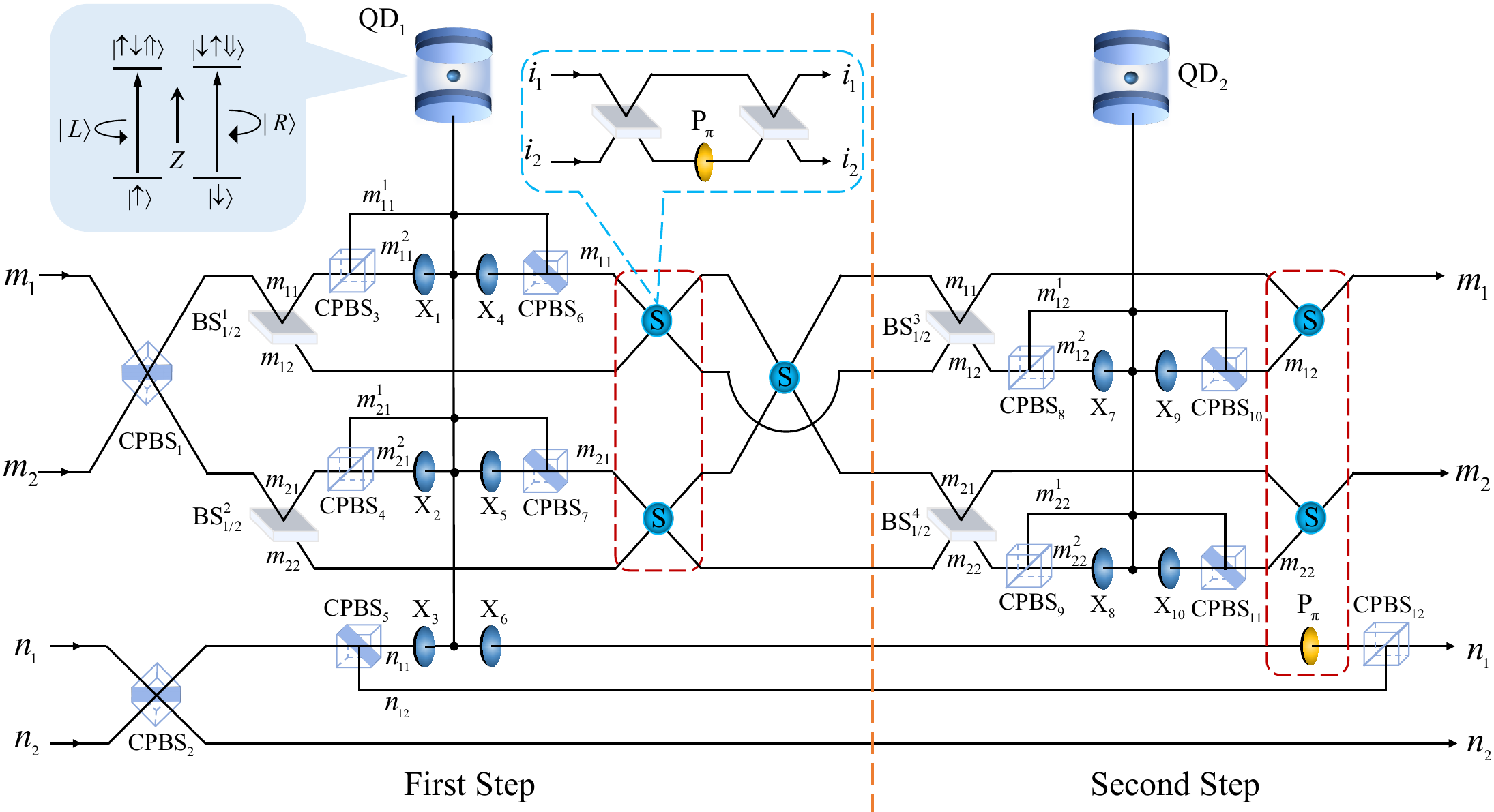}
		\caption{Schematic diagram of the deterministic  16D CSUM gate assisted  by the one-sided microcavity-QD system. $\rm X$$_{k} (k=1,2,...6)$ represents  qubit-flip function on $P$ photon, i.e., $\sigma_{x}^{P} = \vert R\rangle_{P}\langle L|+|L\rangle_{P}\langle R\vert (P=M,N)$.
			P$_\pi$ denotes the phase shifter, performing the transformation, i.e.,
			$|R\rangle \, (|L\rangle) \rightarrow -|R\rangle \, (-|L\rangle)$.
		}\label{fig2}
	\end{center}
\end{figure*}

As illustrated top left in Figure \ref{fig2}, the microcavity-QD system, which consists of a singly charged In(Ga)As QD embedded within an optically resonant single-sided microcavity. The four-level emitter configuration comprises ground states $\vert\downarrow\rangle$ and $\vert\uparrow\rangle$.
This configuration additionally integrates two optically excited $X^{-}$ trion states, labeled as $\vert \uparrow\downarrow\Uparrow\rangle$ and $\vert \downarrow\uparrow\Downarrow\rangle$.
As governed by selection rules, the QD spin states $\vert\downarrow\rangle$ and $\vert\uparrow\rangle$ couple to left- ($L$) and right- ($R$) circularly polarized photons, respectively. This coupling facilitates optical transitions to the respective trion states.
A $ \vert R\rangle $- or $ \vert L\rangle $-polarized photon at frequency \( \omega \), which enters via the input mode \( \hat{a}_{\text{in},\omega}^{\dag} \), interacts with the microcavity-QD system and propagates to the output mode \( \hat{a}_{\text{out},\omega}^{\dag} \). 
Under adiabatic cavity-field evolution under negligible QD excitation, the quantum state-dependent reflection coefficient $r_e$ ($e = 0, 1$) is
\begin{eqnarray} \label{eq2}
	r_{e}(\omega)&=&1-\frac{\kappa f}{eg^{2}+[-(\omega-\omega_{c})i
		+\frac{\kappa}{2}+\frac{\kappa_{s}}{2}]f},
\end{eqnarray}
here, $e=1$ describes the interaction in the coupled microcavity-QD system with the polarized photon, whereas $e=0$ corresponds to the interaction in the uncoupled microcavity configuration. The parameter $g$ denotes the cavity-$X^{-}$-trion coupling strength. The parameter $\kappa$ is the directional coupling rate, governing energy transfer dynamics within the system. $\kappa_s$ corresponds to the cavity-side leakage rate, and $\gamma$ represents the spontaneous emission rate of the trion.
The  $f$ is  $\gamma/2-(\omega - \omega_{X^-})i  $, in which  $\omega_{X^-}$ and $\omega_c$ are the transition frequency of the QD and  the cavity resonance frequency, respectively.
For simplicity, we assume $\omega_{X^-} = \omega_c$ (resonant condition). Under realistic conditions, the state-selective reflection coefficient for the input photon interacting with the one-sided microcavity-QD system is expressed as
\begin{eqnarray} \label{eq3}
	&&|L\rangle|\uparrow\rangle\rightarrow r_{1}|L\rangle|\uparrow\rangle,  \quad |L\rangle|\downarrow\rangle\rightarrow r_{0}|L\rangle|\downarrow\rangle,\nonumber\\
	&&|R\rangle|\downarrow\rangle\rightarrow r_{1}|R\rangle|\downarrow\rangle,  \quad |R\rangle|\uparrow\rangle\rightarrow r_{0}|R\rangle|\uparrow\rangle.
\end{eqnarray}
In the condition $\kappa\gg\kappa_{s}$ and $g^{2}\gg\kappa \gamma$, the coupling and uncoupling reflection coefficients become $r\approx1$ and $r_{0}\approx-1$, leading to the rules
\begin{eqnarray} \label{eq4}
	&&|L\rangle|\uparrow\rangle\rightarrow |L\rangle|\uparrow\rangle,  \quad |L\rangle|\downarrow\rangle\rightarrow -|L\rangle|\downarrow\rangle,\nonumber\\
	&&|R\rangle|\downarrow\rangle\rightarrow |R\rangle|\downarrow\rangle,  \quad |R\rangle|\uparrow\rangle\rightarrow -|R\rangle|\uparrow\rangle.
\end{eqnarray}

Now we introduce the second protocol for  deterministic  16D CSUM gate, which is assisted by two one-sided microcavity-QD systems.
The deterministic 16D CSUM gate is completed with two steps. The first step performs the 4D qudit-flip operation for the 4D target qudit, while the second step serves as a spatial-mode coupler, merging the previously separated four spatial modes into two.
Under these two basic steps, the second protocol is readily implemented.


The quantum circuit for the first step of the deterministic 16D CSUM gate is illustrated in the left part of Figure \ref{fig2}.
The initial states of the two photons, labeled $M$ and $N$, in both polarization and spatial DoFs, are $|\Psi_0\rangle$.
The auxiliary electron spin of the QD$_{j}$-microcavity system is initialized as $|+\rangle_{j} = \frac{1}{\sqrt{2}}(|\uparrow\rangle + |\downarrow\rangle)_{j}$, where $j = 1, 2$ (i.e., QD$_{1}$ of the first step and QD$_{2}$ of the second step). Thus, the initial states of the entire system can be expressed as $|\Phi_0\rangle = |\Psi_0\rangle\otimes |+\rangle_{1}\otimes |+\rangle_{2}$.

Foremostly, similar to the first step of the first protocol, two photons $M$ and $N$ in spatial modes $m_{1}$, $m_{2}$, $n_{1}$, and $n_{2}$ pass through the CPBS$_{1}$ and CPBS$_{2}$, resulting in $|\Phi_1\rangle=|\Psi_1\rangle\otimes|+\rangle_{1}\otimes |+\rangle_{2}$, where $|\Psi_1\rangle$ in Equation (\ref{eq8}).
After that, photon $M$ in spatial mode $m_{1}$ or $m_{2}$ passes through BS$_{1/2}^{1}$ or BS$_{1/2}^{2}$, which apply the operation to two spatial modes of the photon, i.e., $|m_{1}\rangle = (|m_{11}\rangle + |m_{12}\rangle)/\sqrt{2}$ and $|m_{2}\rangle = (|m_{21}\rangle + |m_{22}\rangle)/\sqrt{2}$. Then
the spatial mode of photon $N$ ($M$) through CPBS$_{5}$ (CPBS$_{3}$ or CPBS$_{4}$) is split into two polarization-depended spatial modes.
Subsequently, the $|L\rangle_{M}|L\rangle_{N}$ polarization states of two photons in spatial modes $m_{11}^{1}$ (or $m_{21}^{1}$), and $n_{11}$ pass
directly interact with the microcavity-QD$_{1}$ system via the rules in  Equation (\ref{eq4}), while their $|R\rangle_{M}|R\rangle_{N}$ states in spatial modes $m_{11}^{2}$ (or $m_{21}^{2}$), and $n_{12}$ pass
sequentially through  X$_{1}$ (or X$_{2}$), X$_{3}$ $\rightarrow$  microcavity-QD$_{1}$ system$\rightarrow$ X$_{4}$ (or X$_{5}$), X$_{6}$.
Here, the operator $\rm X_{k}$ ($k=1,2,...,6$) is a half-wave plate executing a qubit-flip function on the $P$ photon, where $\sigma_{x}^{P} = \vert R\rangle_{P}\langle L| + |L\rangle_{P}\langle R|$ for $P = M, N$.
Finally, CPBS$_{6}$ and CPBS$_{7}$ combine the four spatial modes of photon $M$ separated by CPBS$_{3}$ and CPBS$_{4}$. The hybrid state of two photons and electron spin is represented as
\begin{eqnarray} \label{eq12}
	|\Phi_2\rangle &=&
	[\alpha_1 |LL\rangle_{MN} \otimes (\alpha_2 |m_{12} n_{12}\rangle + \beta_2 |m_{12} n_2\rangle+ \gamma_2 |m_{22} n_{12}\rangle+ \delta_2 |m_{22} n_2\rangle) \nonumber\\
	&& + \beta_{1} |LR\rangle_{MN}\otimes (\alpha_2 |m_{12} n_2\rangle + \beta_2 |m_{11} n_{11}\rangle + \gamma_2 |m_{22} n_2\rangle+ \delta_2 |m_{21} n_{11}\rangle) \nonumber\\
	&&+ \gamma_{1} |RL\rangle_{MN} \otimes (\alpha_2 |m_{22} n_{12}\rangle + \beta_2 |m_{22} n_2\rangle+ \gamma_2 |m_{12} n_{12}\rangle + \delta_2 |m_{12} n_2\rangle)\nonumber\\
	&& + \delta_{1} |RR\rangle_{MN} \otimes (\alpha_2 |m_{22} n_2\rangle + \beta_2 |m_{21} n_{11}\rangle+ \gamma_2 |m_{12} n_2\rangle+ \delta_2 |m_{11} n_{11}\rangle)]\otimes|+\rangle_{1}\otimes |+\rangle_{2}\nonumber\\
	&&+[\alpha_1 |LL\rangle_{MN} \otimes (\alpha_2 |m_{11} n_{12}\rangle + \beta_2 |m_{11} n_2\rangle+ \gamma_2 |m_{21} n_{12}\rangle + \delta_2 |m_{21} n_2\rangle)\nonumber\\
	&&  + \beta_{1} |LR\rangle_{MN}\otimes (\alpha_2 |m_{11} n_2\rangle + \beta_2 |m_{12} n_{11}\rangle+ \gamma_2 |m_{21} n_2\rangle+ \delta_2 |m_{22} n_{11}\rangle)\nonumber\\
	&& + \gamma_{1} |RL\rangle_{MN} \otimes (\alpha_2 |m_{21} n_{12}\rangle + \beta_2 |m_{21} n_2\rangle + \gamma_2 |m_{11} n_{12}\rangle + \delta_2 |m_{11} n_2\rangle)\nonumber\\
	&&   + \delta_{1} |RR\rangle_{MN} \otimes (\alpha_2 |m_{21} n_2\rangle + \beta_2 |m_{22} n_{11}\rangle+ \gamma_2 |m_{11} n_2\rangle + \delta_2 |m_{12} n_{11}\rangle)]\otimes|-\rangle_{1}\otimes |+\rangle_{2},
\end{eqnarray}
$|-\rangle_{j} = \frac{1}{\sqrt{2}}(|\uparrow\rangle - |\downarrow\rangle)_{j} (j=1)$.
Then the spin state of the microcavity-QD$_{1}$ system is measured with the basis $\{|+\rangle_{1},|-\rangle_{1}\}$. If the spin is found in state $|+\rangle_{1}$, the  state
\begin{eqnarray} \label{eq13}
	|\Phi_3\rangle &=&
	[\alpha_1 |LL\rangle_{MN} \otimes (\alpha_2 |m_{12} n_{12}\rangle + \beta_2 |m_{12} n_2\rangle+ \gamma_2 |m_{22} n_{12}+ \delta_2 |m_{22} n_2\rangle)\rangle  \nonumber\\
	&&+ \beta_{1} |LR\rangle_{MN}\otimes (\alpha_2 |m_{12} n_2\rangle + \beta_2 |m_{11} n_{11}\rangle + \gamma_2 |m_{22} n_2\rangle+ \delta_2 |m_{21} n_{11}\rangle) \nonumber\\
	&&+ \gamma_{1} |RL\rangle_{MN} \otimes (\alpha_2 |m_{22} n_{12}\rangle+ \beta_2 |m_{22} n_2\rangle + \gamma_2 |m_{12} n_{12}\rangle + \delta_2 |m_{12} n_2\rangle)\nonumber\\
	&&  + \delta_{1} |RR\rangle_{MN} \otimes (\alpha_2 |m_{22} n_2\rangle + \beta_2 |m_{21} n_{11}\rangle+ \gamma_2 |m_{12} n_2\rangle + \delta_2 |m_{11} n_{11}\rangle)]\otimes |+\rangle_{2}
\end{eqnarray}
is obtained. Otherwise, if the spin is in state $|-\rangle_{1}$, the related feed-forward operation with two swap (S) gates fulfilling by two BSs$_{1/2}$ and a phase shifter $\sigma_{z}$, executing the unitary operator $\sigma^{S} = |m_{q1}\rangle\langle m_{q2}| + |m_{q2}\rangle\langle m_{q1}|, (q=1,2)$ on two spatial modes, is applied to photon $M$ to get the desired state $|\Phi_3\rangle$ in  Equation (\ref{eq13}).
Finally, photon $M$ in spatial modes $m_{11}$ and $m_{21}$ undergoes another S to exchange two spatial modes, leading to
\begin{eqnarray} \label{eq14}
	|\Phi_4\rangle &=&
	[\alpha_1 |LL\rangle_{MN} \otimes (\alpha_2 |m_{12} n_{12}\rangle + \beta_2 |m_{12} n_2\rangle+ \gamma_2 |m_{22} n_{12}\rangle+ \delta_2 |m_{22} n_2\rangle) \nonumber\\
	&&  + \beta_{1} |LR\rangle_{MN}\otimes (\alpha_2 |m_{12} n_2\rangle + \beta_2 |m_{11} n_{11}\rangle + \gamma_2 |m_{22} n_2\rangle+ \delta_2 |m_{21} n_{11}\rangle) \nonumber\\
	&&+ \gamma_{1} |RL\rangle_{MN} \otimes (\alpha_2 |m_{22} n_{12}\rangle+ \beta_2 |m_{22} n_2\rangle + \gamma_2 |m_{12} n_{12}\rangle + \delta_2 |m_{12} n_2\rangle)\nonumber\\
	&&  + \delta_{1} |RR\rangle_{MN}  \otimes (\alpha_2 |m_{22} n_2\rangle + \beta_2 |m_{21} n_{11}\rangle+ \gamma_2 |m_{12} n_2\rangle + \delta_2 |m_{11} n_{11}\rangle)]\otimes |+\rangle_{2}.
\end{eqnarray}
Upon completing first step, the target qudit reaches the desired state.



The quantum circuit for the second step of the deterministic  16D CSUM  gate is illustrated in the right part of Figure \ref{fig2}.
Firstly, photon $M$ passes through BS$_{1/2}^{3}$ or BS$_{1/2}^{4}$, namely,
$|m_{q1}\rangle = (|m_{q1}\rangle + |m_{q2}\rangle)/\sqrt{2}$ and $|m_{q2}\rangle = (|m_{q1}\rangle - |m_{q2}\rangle)/\sqrt{2}, (q=1,2)$.
Subsequently, photon $M$ in spatial modes $m_{12}$ and $m_{22}$ undergoes the processes, that is, passing through CPBS$_{8}$ (CPBS$_{9}$),
interacting with the microcavity-QD$_{2}$ system, and converging at CPBS$_{10}$ (CPBS$_{11}$), the same as the former microcavity-QD$_{1}$ system, thereby resulting in the state $|\Phi_4\rangle$ transformed into
\begin{eqnarray} \label{eq15}
	|\Phi_5\rangle&=& [\alpha_1|LL\rangle_{MN}\otimes(\alpha_2|m_{11}n_{12}\rangle+\beta_2|m_{11}n_2\rangle+\gamma_2|m_{21}n_{12}\rangle+\delta_2|m_{21}n_2\rangle)\nonumber\\
	&&+\beta_{1}|LR\rangle_{MN}\otimes(\alpha_2|m_{11}n_2\rangle+\beta_2|m_{21}n_{11}\rangle+\gamma_2|m_{21}n_2\rangle+\delta_2|m_{11}n_{11}\rangle)\nonumber\\
	&&+\gamma_{1}|RL\rangle_{MN}\otimes(\alpha_2|m_{21}n_{12}\rangle+\beta_2|m_{21}n_2\rangle+\gamma_2|m_{11}n_{12}\rangle+\delta_2|m_{11}n_2\rangle)\nonumber\\
	&&+\delta_{1}|RR\rangle_{MN} \otimes(\alpha_2|m_{21}n_2\rangle+\beta_2||m_{11}n_{11}\rangle+\gamma_2|m_{11}n_2\rangle+\delta_2|m_{21}n_{11}\rangle)]\otimes|+\rangle_{2}\nonumber\\
	&&-[\alpha_1|LL\rangle_{MN}\otimes(\alpha_2|m_{12}n_{12}\rangle+\beta_2|m_{12}n_2+\gamma_2|m_{22}n_{12}\rangle+\delta_2|m_{22}n_2\rangle)\rangle\nonumber\\
	&&+\beta_{1}|LR\rangle_{MN}\otimes(\alpha_2|m_{12}n_2\rangle-\beta_2|m_{22}n_1\rangle+\gamma_2|m_{22}n_2\rangle-\delta_2|m_{12}n_{11}\rangle)\nonumber\\
	&&+\gamma_{1}|RL\rangle_{MN}\otimes(\alpha_2|m_{22}n_1\rangle+\beta_2|m_{22}n_2\rangle+\gamma_2|m_{12}n_{11}\rangle+\delta_2|m_{12}n_2\rangle)\nonumber\\
	&&+\delta_{1}|RR\rangle_{MN} \otimes(\alpha_2|m_{22}n_2\rangle-\beta_2|m_{12}n_{12}\rangle+\gamma_2|m_{12}n_2\rangle-\delta_2|m_{22}n_{12}\rangle)]\otimes|-\rangle_{2}.
\end{eqnarray}
Then the spin state of the microcavity-QD$_{2}$ system is measured with the basis $\{|+\rangle_{2},|-\rangle_{2}\}$.
If the spin is found in state $|+\rangle_{2}$, the state
\begin{eqnarray} \label{eq16}
	|\Phi_6\rangle&=& \alpha_1|LL\rangle_{MN}\otimes(\alpha_2|m_{11}n_{12}\rangle+\beta_2|m_{11}n_2\rangle+\gamma_2|m_{21}n_{12}\rangle+\delta_2|m_{21}n_2\rangle)\nonumber\\
	&&+\beta_{1}|LR\rangle_{MN}\otimes(\alpha_2|m_{11}n_2\rangle+\beta_2|m_{21}n_{11}\rangle+\gamma_2|m_{21}n_2\rangle+\delta_2|m_{11}n_{11}\rangle)\nonumber\\
	&&+\gamma_{1}|RL\rangle_{MN}\otimes(\alpha_2|m_{21}n_{12}\rangle+\beta_2|m_{21}n_2\rangle+\gamma_2|m_{11}n_{12}\rangle+\delta_2|m_{11}n_2\rangle)\nonumber\\
	&&+\delta_{1}|RR\rangle_{MN} \otimes(\alpha_2|m_{21}n_2\rangle+\beta_2||m_{11}n_{11}\rangle+\gamma_2|m_{11}n_2\rangle+\delta_2|m_{21}n_{11}\rangle)
\end{eqnarray}
is obtained. Apparently, the  deterministic  16D CSUM gate is successfully constructed.
On the contrary, if the spin is in state $|-\rangle_{2}$, the related feed-forward operation assisted by two S gates on two pairs of spatial modes of photon $M$,
$m_{12}$ and $m_{11}$, $m_{22}$ and $m_{21}$, is applied to photon $M$ to get the desired state $|\Psi_6\rangle$ in  Equation (\ref{eq16}).
Thus, two spatial modes $m_{11}$ and $m_{12}$ ($m_{21}$ and $m_{22}$) are coupled to one spatial mode $m_{1}$ ($m_{2}$). Additionally, photon $N$ in spatial mode $n_{11}$ should pass through a $P_{\pi}$ operation. Finally, two spatial modes $n_{11}$ and $n_{12}$ of photon $N$ converge into spatial modes $n_1$ through CPBS$_{12}$. The entire system state is transformed into
\begin{eqnarray} \label{eq17}
	|\Phi_7\rangle&=& \alpha_1|LL\rangle_{MN}\otimes(\alpha_2|m_{1}n_1\rangle+\beta_2|m_{1}n_2\rangle+\gamma_2|m_{2}n_1\rangle+\delta_2|m_{2}n_2\rangle)\nonumber\\
	&&+\beta_{1}|LR\rangle_{MN}\otimes(\alpha_2|m_{1}n_2\rangle+\beta_2|m_{2}n_1\rangle+\gamma_2|m_{2}n_2\rangle+\delta_2|m_{1}n_1\rangle)\nonumber\\
	&&+\gamma_{1}|RL\rangle_{MN}\otimes(\alpha_2|m_{2}n_1\rangle+\beta_2|m_{2}n_2\rangle+\gamma_2|m_{1}n_1\rangle+\delta_2|m_{1}n_2\rangle)\nonumber\\
	&&+\delta_{1}|RR\rangle_{MN} \otimes(\alpha_2|m_{2}n_2\rangle+\beta_2||m_{1}n_1\rangle+\gamma_2|m_{1}n_2\rangle+\delta_2|m_{2}n_1\rangle)\nonumber\\
	&=&
	\alpha_1|\bar{0}\rangle_{c}\otimes(\alpha_2|\bar{0}\rangle_{t}+\beta_2|\bar{1}\rangle_{t}
	+\gamma_2|\bar{2}\rangle_{t}+\delta_2|\bar{3}\rangle_{t})\nonumber\\
	&&+\beta_{1}|\bar{1}\rangle_{c}\otimes(\alpha_2|\bar{1}\rangle_{t}+\beta_2|\bar{2}\rangle_{t}+\gamma_2|\bar{3}\rangle_{t}+\delta_2|\bar{0}\rangle_{t})\nonumber\\
	&&
	+\gamma_{1}|\bar{2}\rangle_{c}\otimes(\alpha_2|\bar{2}\rangle_{t}+\beta_2|\bar{3}\rangle_{t}
	+\gamma_2|\bar{0}\rangle_{t}+\delta_2|\bar{1}\rangle_{t})\nonumber\\
	&&+\delta_{1}|\bar{3}\rangle_{c} \otimes(\alpha_2|\bar{3}\rangle_{t}+\beta_2|\bar{0}\rangle_{t}
	+\gamma_2|\bar{1}\rangle_{t}+\delta_2|\bar{2}\rangle_{t}).
\end{eqnarray}
We have proposed the second protocol to implement the deterministic  16D CSUM gate assisted by the photon scattering property in the two QD-microcavity systems.

\section{The Error-heralded deterministic 16D CSUM gate}  \label{sec4}

\begin{figure*}[t]
	\centering
	\begin{center}
		\centering
		\includegraphics[width=1\linewidth]{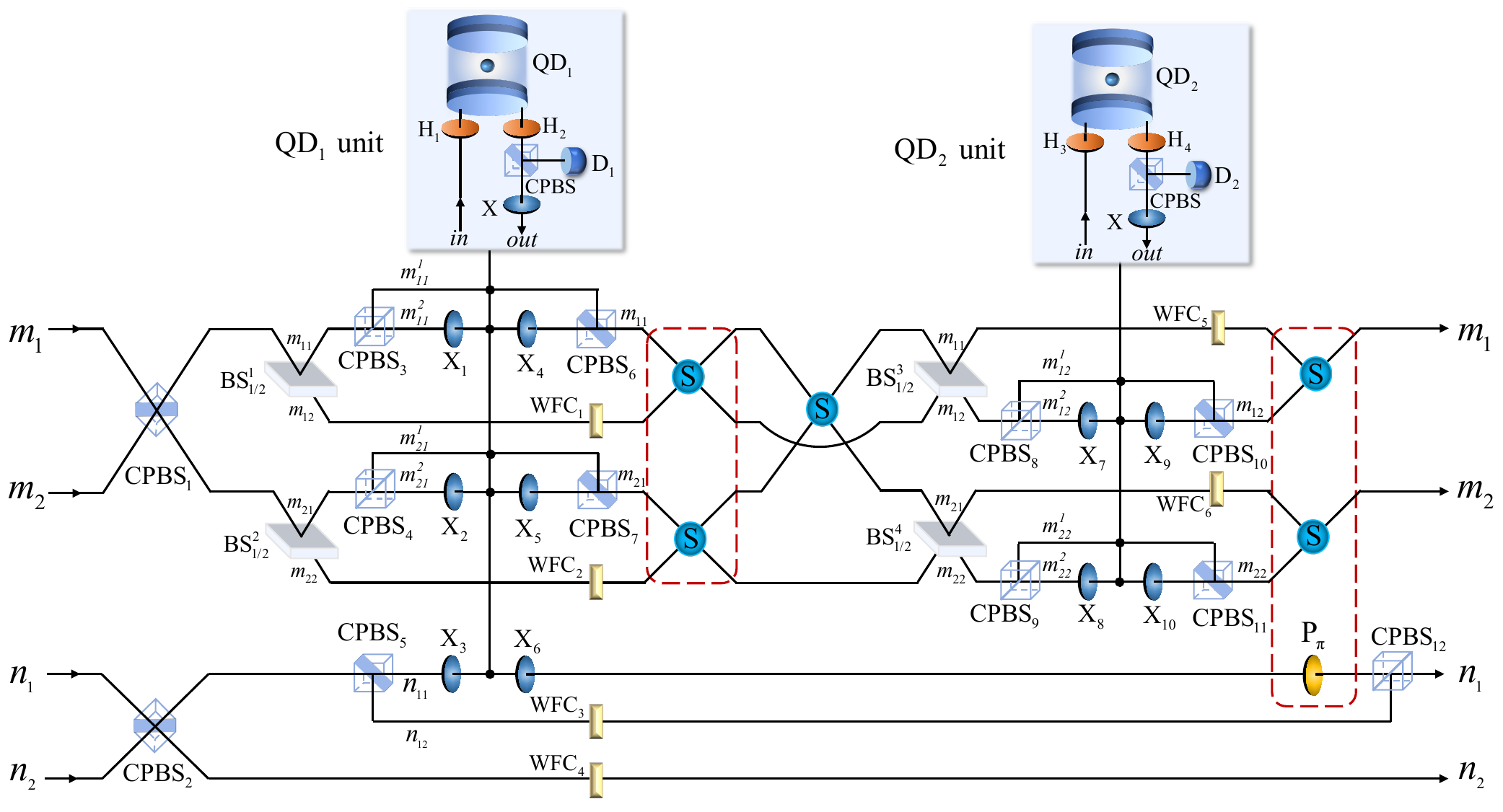}
		\caption{Schematic diagram of the error-heralded  16D CSUM gate.  WFC is a wave-form corrector that executes the coefficient variation \( \vert R \rangle \) (\( \vert L \rangle \)) $\rightarrow$ \( A\vert R \rangle \)(\( \vert L \rangle \)). Half-wave plate H$_{i}$ ($i = 1,2,3,4$)
			performs the Hadamard operation in polarized DoF.}\label{fig3}
	\end{center}
\end{figure*}

The fidelity of deterministic  16D CSUM gate assisted  by the one-sided microcavity-QD system is affected by non-ideal scattering, including defective coupling between the cavity mode and the QD and minor side-cavity leakage, leading to a fidelity less than unity. To address this, we propose the third protocol, i.e., an error-heralded  deterministic 16D CSUM gate shown in Figure \ref{fig3}, where erroneous components are passively filtered by the CPBS, and error prediction is achieved using a single-photon detector (D), resulting in unity fidelity with ignoring photon loss.
Compared with the above CSUM gate in  Equation (\ref{eq17}) formed in Figure \ref{fig2},
the quantum circuit of the error-heralded  deterministic 16D CSUM gate shown in Figure \ref{fig3}, replaces the QD$_{j}$-microcavity system with the error-heralded QD$_{j}$ unit and adds some wave-front correctors (WFCs) to designated spatial modes, where the photon does not interact with the error-heralded QD$_{j}$ unit during each interaction. WFC executes the coefficient variation \( \vert R \rangle \) (\( \vert L \rangle \)) $\rightarrow$ \( A\vert R \rangle \)(\( \vert L \rangle \)). As the $|L\rangle$-polarized photon traverses error-heralded QD$_{j}$ unit, it encounters various optical elements, that is, H$_{1}$ (H$_{3}$) $\rightarrow$ microcavity-QD$_{j}$ system via the rules in  Equation (\ref{eq3})$\rightarrow$ H$_{2}$ (H$_{4})$.
Here, the operator \( H_i \) (for \( i = 1, 2, 3, 4 \)) denotes a half-wave plate that implements the Hadamard transformation on the photon's polarization DoF, such that \( \vert L\rangle \rightarrow (\vert R\rangle - \vert L\rangle)/\sqrt{2} \) and \( \vert R\rangle \rightarrow (\vert R\rangle + \vert L\rangle)/\sqrt{2} \).
As a result of these operations, the system evolves into
\begin{eqnarray}  \label{eq18}
	\vert L\rangle\vert+\rangle\rightarrow A\vert R\rangle\vert-\rangle+B\vert L\rangle\vert +\rangle,
\end{eqnarray}
where \( A \) and \( B \) are expressed as \( (r_1 - r_0)/2 \) and \( (r_1 + r_0)/2 \), respectively. As governed by Equation (\ref{eq18}), an input \( |L\rangle \)-polarized photon generates two distinct output states. Specifically, detector D\(_j\) detects a photon in the \( |L\rangle \) output state upon reflection by the CPBS, indicating a fault. However, the spin state of the microcavity-QD\(_j\) system retains its integrity and remains reusable in subsequent cycles. The polarization-dependent selection imposed by the CPBS suppresses errors stemming from imperfect photon-spin interactions within the microcavity-QD\(_j\) system. Conversely, a photon in the \( |R\rangle \) output state propagates directly through the CPBS and subsequently passes through an X-polarization converter, yielding the final desired outcome
\begin{eqnarray}  \label{eq19}
	\vert L\rangle\vert+\rangle\rightarrow A\vert L\rangle\vert-\rangle.
\end{eqnarray}
Thus, assuming identical initial states for both gates, the states in each expression of the error-heralded  deterministic 16D CSUM gate, corresponding to the deterministic 16D CSUM gate assisted  by the one-sided microcavity-QD system, are
\begin{eqnarray}  \label{eq20}
	|\rm CSUM^{'}\rangle = A^{3} |\rm \Phi_{7}\rangle.
\end{eqnarray}

\section{Discussion on the fidelity and efficiency of each protocol  }\label{sec5}

The first protocol relies solely on linear optical elements, including CPBSs and BSs. Experimentally, CPBS may be substituted with calcite beam displacer (BD), as both optical components direct photons into different spatial modes based on their polarization states. Thus, upon the photon passes through the first CPBS, misalignments in mirror mounts ($\phi$) and the polarization extinction ratio ($p$) result in deviations of the actual quantum state from the ideal input state. Consequently, the mode transformation matrix, applied to the two creation operators and influenced by $\phi$ and $p$, is expressed as \cite{PhysRevA.105.032607}
\begin{equation} \label{eqf4}
	\hat{U}^{1}_{\mathrm{CPBS}}=\hat{U}_{p}\hat{U}_\phi=
	\begin{bmatrix}
		\frac{1}{\sqrt{1+p}} & \frac{\sqrt{p}}{\sqrt{1+p}} \\
		\frac{-\sqrt{p}^*}{\sqrt{1+p}} & \frac{1}{\sqrt{1+p}}
	\end{bmatrix}
	\begin{bmatrix}
		\cos\phi & \sin\phi \\
		-\sin\phi & \cos\phi
	\end{bmatrix},\end{equation}
$\sqrt{p}^*$ denotes the
conjugate complex of $\sqrt{p}$.
Assume that the input state is
$|R\rangle$ or $|L\rangle$, the fidelity is given by
\begin{align}\label{eqf6}
	\mathcal{F}_R &= \left| \langle \varphi_{\text{real}} | \varphi_{\text{ideal}} \rangle \right|^2 = \frac{1}{1 + p} \left|   \cos \phi - \sqrt{r} \sin \phi  \right|^2,\nonumber\\
	\mathcal{F}_L &= \left| \langle \varphi_{\text{real}} | \varphi_{\text{ideal}} \rangle \right|^2 = \frac{1}{1 + p}\left|-\sqrt{r}^* \sin\phi + \cos\phi\right|^{2}.
\end{align}
By analyzing Equation (\ref{eqf6}), it is determined that \( \mathcal{F}_R = \mathcal{F}_L \). Consequently, the fidelity of the first CPBS, denoted as \( \mathcal{F}^{1}_{\mathrm{CPBS}} \), is given by \( \frac{1}{1 + p} \left| \cos \phi - \sqrt{p} \sin \phi \right|^2 \).

\begin{figure*}[t]
	\centering
	\begin{center}
		\centering
		\includegraphics[width=1\linewidth]{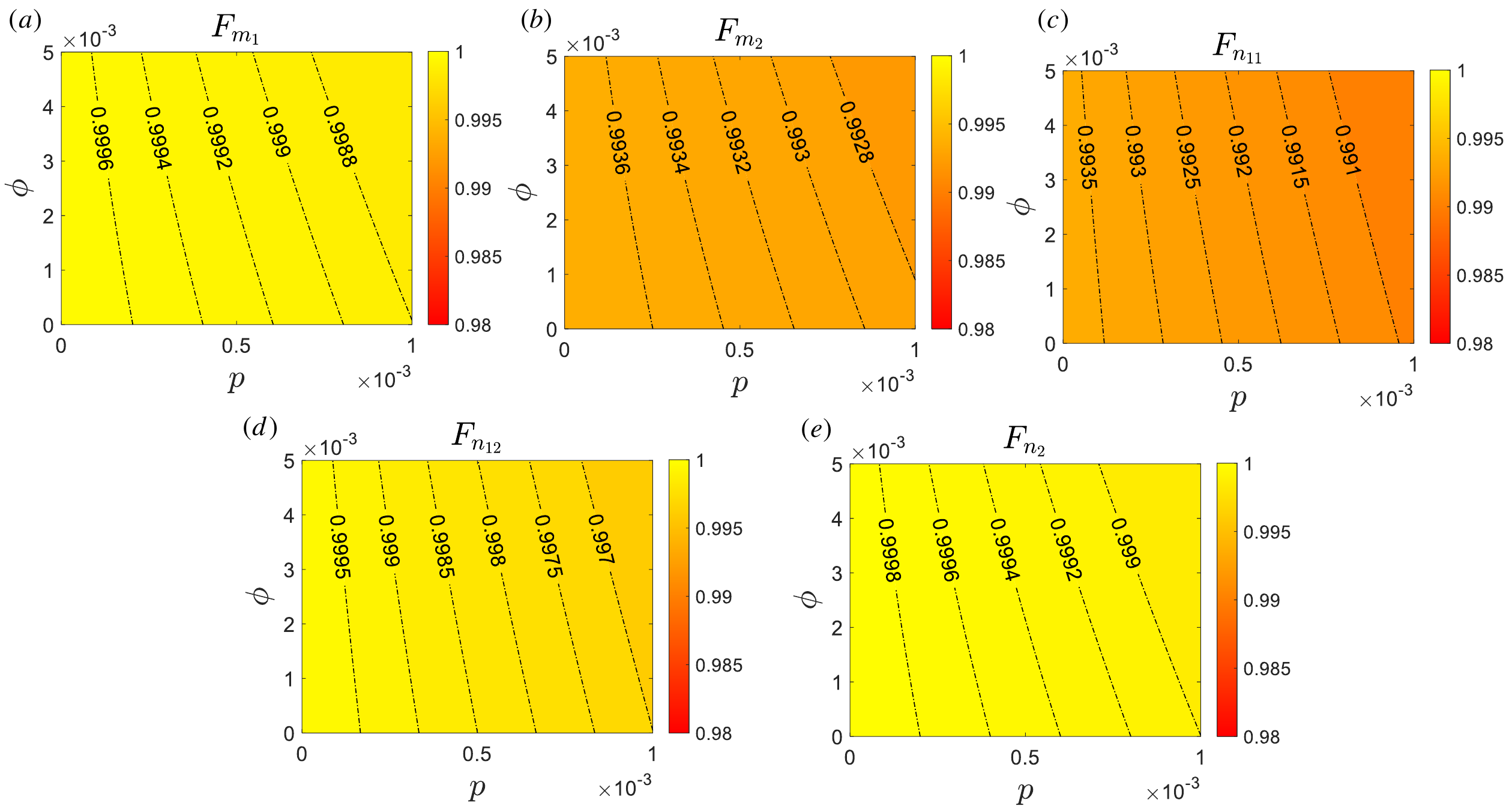}
		\caption{The fidelity of (a) photon $M$ in path $m_{1}$ (b) photon $M$ in path $m_{2}$ (c) photon $N$ in path $n_{11}$ (d) photon $N$ in path $n_{12}$ (e) photon $N$ in path $n_{2}$ vs  the deviation in mirror mounts $\phi$  and the polarization extinction ratio $p$ under realistic conditions \( \Delta = \pi/36 \) and \( \xi = 0.02 \).}\label{fig4}
	\end{center}
\end{figure*}

\begin{figure}[t]
	\centering
	\begin{center}
		\centering
		\includegraphics[width=0.75\linewidth]{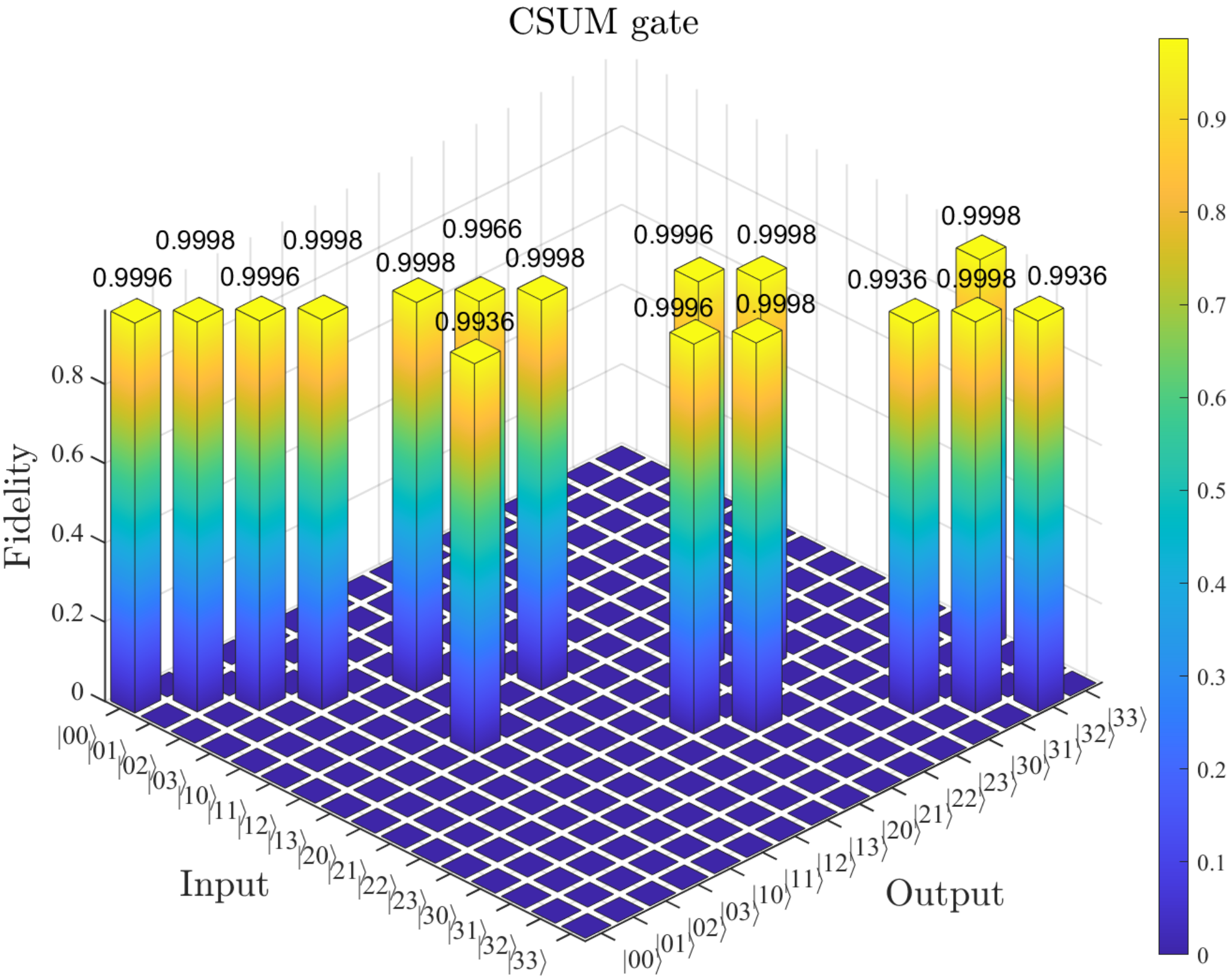}
		\caption{The fidelity of the first protocol  under realistic conditions \( \Delta = \pi/36 \), \( \xi = 0.02 \), $\phi=0.001$ and $p=0.0001$.}\label{fig5}
	\end{center}
\end{figure}

For simplicity, it is assumed that the second CPBS and the first CPBS exhibit an identical extinction ratio. Ideally, the unitary mode transformation matrix of the second CPBS constitutes the inverse of the transformation matrix of the first CPBS rotated by 90$^{\circ}$, i.e.,
\begin{equation}\label{eqf7}
	\hat{U}'_{p}=\frac{1}{\sqrt{1+p}}{
		\begin{bmatrix}
			1 & \sqrt{p}^{*} \\
			-\sqrt{p} & 1
	\end{bmatrix}}.\end{equation}
In this step, the angle of deviation of the mirror mounts is not reconsidered, as the first CPBS and second CPBS are typically assembled with a consistent orientation. Consequently, by the definition of fidelity
\begin{equation} \label{eqf8}
	\mathcal{F}^{2}_{\mathrm{CPBS}}=\mathcal{F}_R'=\mathcal{F}_L'=|\langle\varphi_\mathrm{real}|\varphi_\mathrm{ideal}\rangle|^2=\frac{1}{1+p}.
\end{equation}

\begin{figure*}
	\centering
	\begin{center}
		\centering
		\includegraphics[width=0.8\linewidth]{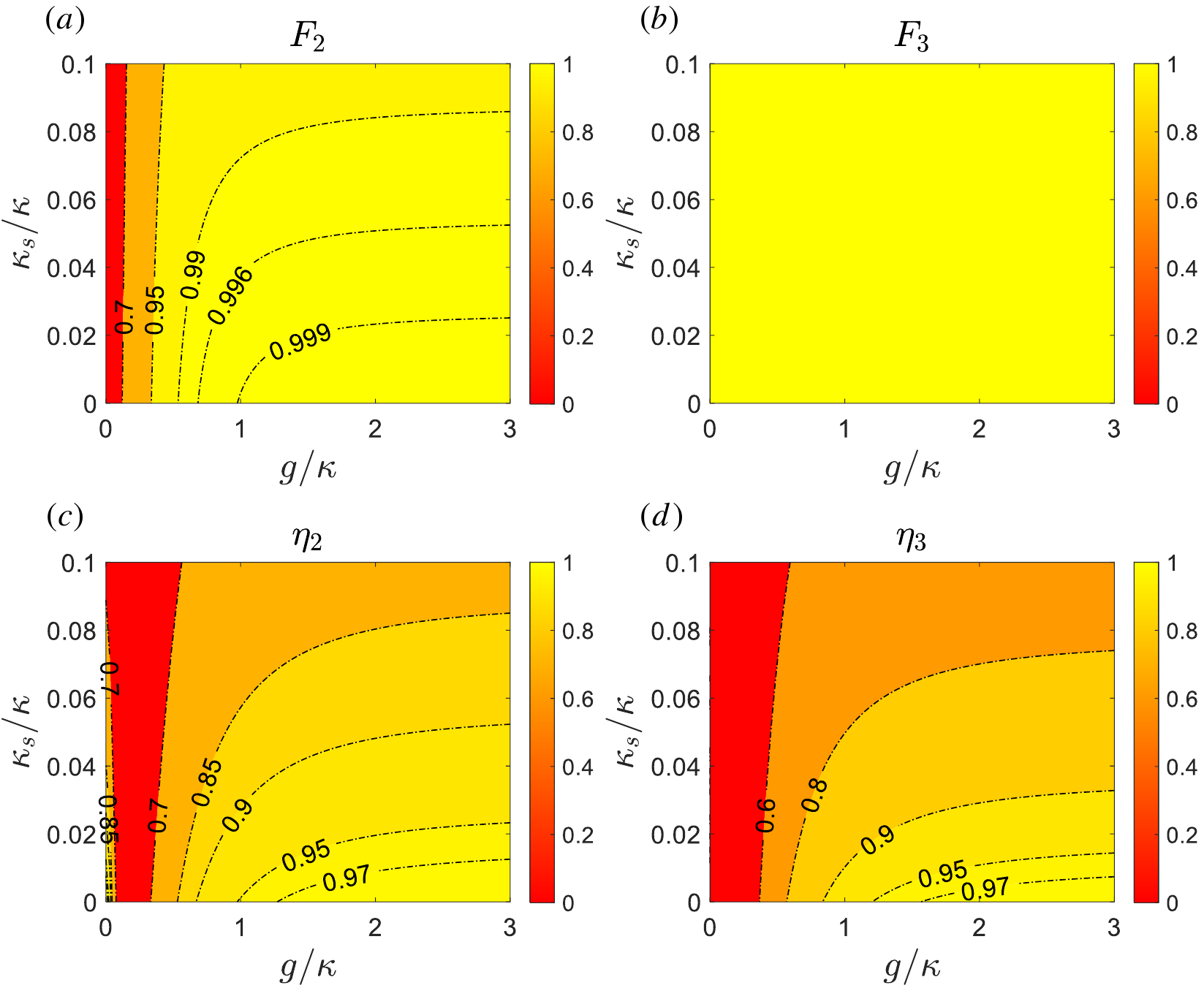}
		\caption{(a) the fidelity $F_{2}$  of the second protocol (b) the fidelity $F_{3}$  of the third protocol (ignoring photon loss) (c) the efficiency $\eta_{2}$ of the second protocol (d) the efficiency $\eta_{3}$  of the third protocol vs  the side leakage rate $\kappa_{s}/\kappa$ and the coupling strength $g/\kappa$  with $\gamma = 0.1\kappa$.}\label{fig6}
	\end{center}
\end{figure*}

The interference network formed by the combination of BSs$_{1/2}$ and BSs$_{1/3}$ essentially constitutes a Mach-Zehnder interferometer (MZI).
If the actual phase difference between the two arms of the MZI is given by $\pi - \Delta$, the unitary transformation matrix describing the internal phase-shifting element is expressed as
\begin{equation}\label{eqf15}
	\hat{U}_{\pi - \Delta} =
	\begin{bmatrix}
		e^{i(\pi - \Delta)} & 0 \\
		0 & 1
	\end{bmatrix}.
\end{equation}
Assuming a small deviation from ideal transmittance in BS$_{1/2}$, the unitary transformation matrix of the BS$_{1/2}$ is
\begin{equation}\label{eqf16}
	\hat{U}_{\text{BS$_{1/2}$}} =
	\frac{1}{\sqrt{\xi^2 + 2\xi + 2}}
	\begin{bmatrix}
		1 + \xi & i \\
		i & 1 + \xi
	\end{bmatrix}.
\end{equation}
where $\xi$ denotes the imperfection in the transmission ratio.
Consequently, the fidelity of BS$_{1/2}$ is
\begin{equation}\label{eqf18}
	\mathcal{F}_{\text{BS$_{1/2}$}} = \left| \langle \varphi_{\text{real}} | \varphi_{\text{ideal}} \rangle \right|^2 = \frac{(\xi + 2)^{2}}{2 \left(\xi^2 + 2\xi + 2 \right)}.
\end{equation}
Then the
unitary transformation matrix for an imperfect MZI becomes
\begin{eqnarray}\label{eqf17}
	\hat{U}_{\text{MZI}} &=& \hat{U}_{\text{BS$_{1/2}$}} \hat{U}_{\pi - \Delta} \hat{U}_{\text{BS$_{1/2}$}}\nonumber\\
	&=&
	\frac{1}{\xi^2 + 2\xi + 2}\begin{bmatrix}
		e^{i(\pi - \Delta)}(1 + \xi)^2 - 1 & i(1 + \xi)(e^{i(\pi - \Delta)} + 1) \\
		i(1 + \xi)(e^{i(\pi - \Delta)} - 1) & -e^{i(\pi - \Delta)} + (1 + \xi)^2
	\end{bmatrix}.\nonumber\\
\end{eqnarray}
The fidelity of MZI is then given by
\begin{equation}\label{eqf18}
	\mathcal{F}_{\text{MZI}} = \left| \langle \varphi_{\text{real}} | \varphi_{\text{ideal}} \rangle \right|^2 = \frac{(1 + \cos(\Delta)) \left[ (1 + \xi)^2 + 1 \right]^2}{2 \left( \xi^2 + 2\xi + 2 \right)^2}.
\end{equation}
The BS\(_{1/3}^{1}\), BS\(_{1/3}^{3}\), and BS\(_{1/3}^{4}\) only serves to balance the losses, and thus its impact on the overall fidelity can be considered negligible.

We calculate the fidelity of the first protocol  based on linear elements encountered by the photon in different paths.
The fidelity of photon $M$ in path $m_{1}$  can be expressed as
\begin{eqnarray}\label{eqf24}
	{F}_{m_{1}} &=&\mathcal{F}^{1}_{\mathrm{CPBS}}\mathcal{F}_{\mathrm{BS_{1/2}}}^{2}\nonumber\\
	&=&\frac{\left|   \cos \phi - \sqrt{r} \sin \phi \right|^{2}}{1 + p} \frac{(\xi + 2)^{4}}{4 \left( \xi^2 + 2\xi + 2 \right)^{2}}.
\end{eqnarray}
The fidelity of photon $M$ in path $m_{2}$  can be expressed as
\begin{eqnarray}\label{eqf24}
	{F}_{m_{2}} &=&\mathcal{F}^{1}_{\mathrm{CPBS}}\mathcal{F}_{\mathrm{MZI}}\nonumber\\
	&=&\frac{\left|   \cos \phi - \sqrt{r} \sin \phi  \right|^{2}}{1 + p} \frac{(1 + \cos(\Delta)) \left[ (1 + \xi)^2 + 1 \right]^2}{2\left( \xi^2 + 2\xi + 2 \right)^{2}}.
\end{eqnarray}
The fidelity of photon $N$ in path $n_{11}$ can be expressed as
\begin{eqnarray}\label{eqf24}
	{F}_{n_{11}} &=&(\mathcal{F}^{1}_{\mathrm{CPBS}})^{2}\mathcal{F}^{2}_{\mathrm{CPBS}}\mathcal{F}_{\mathrm{MZI}}\nonumber\\
	&=&\frac{\left|   \cos \phi - \sqrt{r} \sin \phi \right|^{4}}{(1 + p)^{3}}\frac{(1 + \cos(\Delta)) \left[ (1 + \xi)^2 + 1 \right]^2}{2 \left( \xi^2 + 2\xi + 2 \right)^2}.
\end{eqnarray}
The fidelity of photon $N$ in path $n_{12}$ can be expressed as
\begin{eqnarray}\label{eqf24}
	{F}_{n_{12}} =(\mathcal{F}^{1}_{\mathrm{CPBS}})^{2}\mathcal{F}^{2}_{\mathrm{CPBS}}
	=\frac{\left|   \cos \phi - \sqrt{r} \sin \phi  \right|^{4}}{(1 + p)^{3}}.
\end{eqnarray}
The fidelity of photon $N$ in path $n_{2}$ can be expressed as
\begin{eqnarray}\label{eqf24}
	{F}_{n_{2}} =\mathcal{F}^{1}_{\mathrm{CPBS}}
	=\frac{\left|   \cos \phi - \sqrt{r} \sin \phi  \right|^{2}}{1 + p}.
\end{eqnarray}
Assuming realistic conditions $\Delta = \pi/36$ and $\xi = 0.02$,  the fidelities ${F}_{m_{1}}$, ${F}_{m_{2}}$, ${F}_{n_{11}}$, ${F}_{n_{12}}$, ${F}_{n_{2}}$  are shown in Figures \ref{fig4}(a)-(e), respectively.
When linear optical elements exhibit defects or nonidealities, i.e., delay error $\Delta = \pi/36$, imperfection of
the transmission ratio $\xi = 0.02$,  the deviation in mirror mounts $\phi=0.001$,  and the polarization extinction ratio $p=0.0001$, the fidelity of the first protocol is calculated to be greater than 0.9936 with 16  computational basis  as shown in Figure \ref{fig5}.

\begin{figure}[t]
	\centering
	\begin{center}
		\centering
		\includegraphics[width=0.75\linewidth]{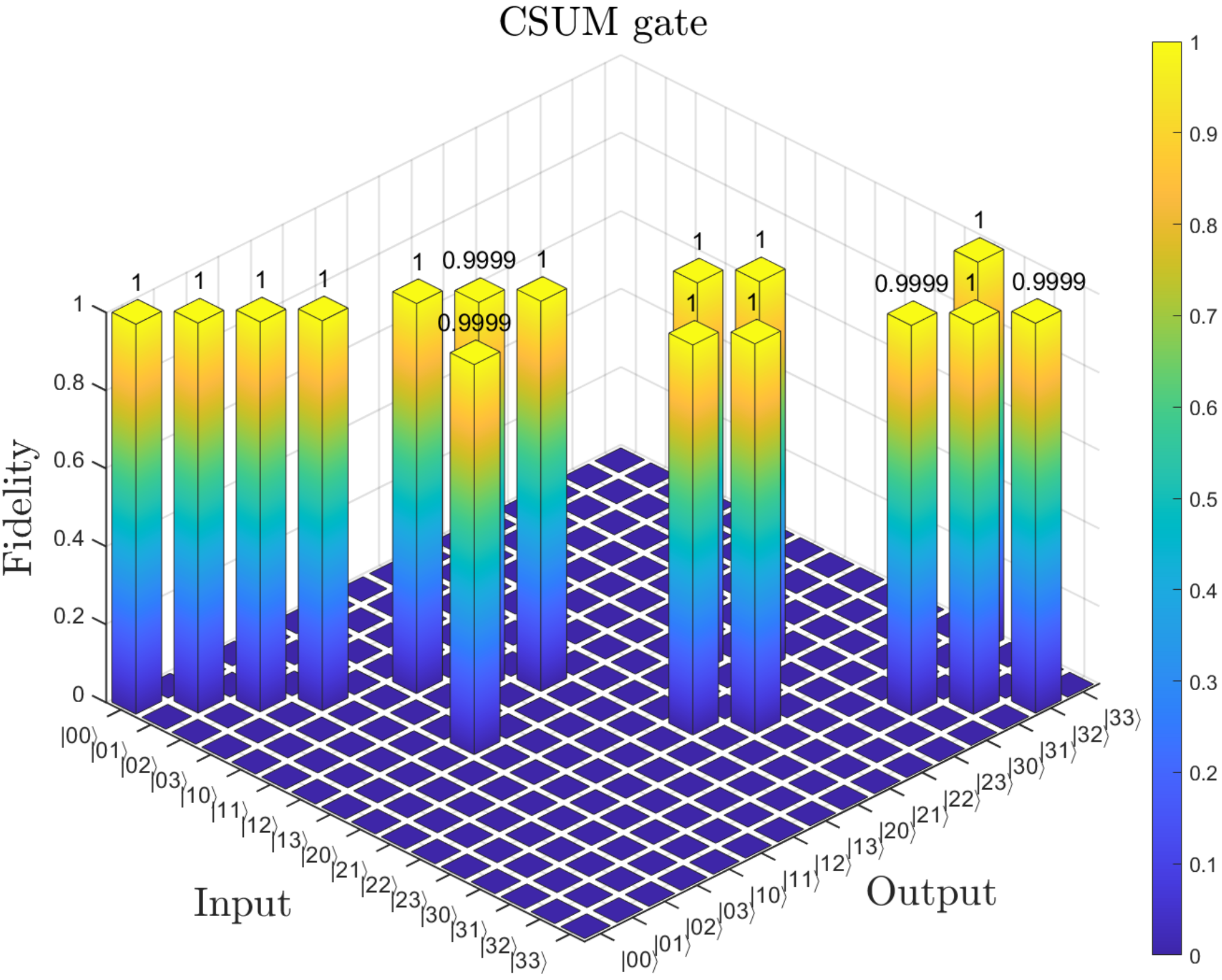}
		\caption{The fidelity of the second protocol with computational basis under the condition $g/\kappa=2.4$, $\kappa_{s}/\kappa=0$ and $\gamma = 0.1\kappa$. }\label{fig7}
	\end{center}
\end{figure}

For the second protocol that present the  deterministic 16D CSUM gate, assuming that the side leakage rate $\kappa_s$ is negligible, the theoretical fidelity can reach unity in perfect microcavity system. However, to present the more realistic scenario, the influence of $\kappa_s$ must be considered. The side leakage rate $\kappa_s$ affects the amplitude of the reflected photon, which, in turn, impacts the fidelity $F_2$ and efficiency $\eta_2$ of the 16D CSUM gate. 
For the error-heralded 16D CSUM gate of the third protocol, the two QD$_{1}$ and QD$_{2}$ units have the error-heralded mechanism during imperfect photon scattering process. Errors arising from non-ideal scattering are filtered by the CPBSs. For example,  the CPBS of the QD$_{j}$ unit passively filters out the incorrect $|L\rangle$ state components while transmitting the useful $|R\rangle$ state components to X. The response of D is solely used to indicate the presence of error.
As a result, the fidelity $F_{3}$ of the error-heralded deterministic 16D CSUM gate shown in Equation (\ref{eq20}), can theoretically approach unity  (ignoring photon loss) as shown in Figure \ref{fig6} (b), but impacts its efficiency $\eta_3$, leading to lower than the efficiency $\eta_2$.
In a word,	
the cavity decay rate ratio $\kappa_s/\kappa$ and the coupling strength ratio $g/\kappa$ significantly influence the efficiency $\eta_2$ and $\eta_3$ of both two protocols and fidelity $F_{2}$ of the second protocol, but not affect the fidelity $F_{3}$ of the third protocol. In practical experiments, a value of $\gamma = 0.1\kappa$ is achievable. Specifically, the analyses of the fidelities and efficiencies reveal that they are improved as $g/\kappa$ increases and $\kappa_s/\kappa$ decreases
in Figure \ref{fig6}.
For instance,  when $g/\kappa$ increases from 1.2 to 2.4, $F_{2}$, $\eta_2$, and $\eta_3$ improve from 0.9995, 0.9670, and 0.9506 to 0.9999, 0.9913, and 0.9870, respectively, while maintaining a constant $\kappa_s/\kappa = 0$. Conversely, when $\kappa_s/\kappa$ increases from 0 to 0.05 while keeping $g/\kappa = 2.4$, $F_{2}$, $\eta_2$, and $\eta_3$ decrease from 0.9999, 0.9913, and 0.9870 to 0.9961, 0.8998, and 0.8508, respectively.

In practical implementations, achieving strong coupling represents a prerequisite for robust-fidelity 16D gate operation. Strong coupling has been experimentally demonstrated in cavity-QD\(_j\) systems, including micropillar structures, microdisk configurations, and QD-nanocavity architectures. In 2011, Hu $et$ $al.$ reported \( g/\kappa \approx 2.4 \) in a micropillar cavity featuring \( \kappa_s/\kappa \approx 0 \) and a quality factor \( Q \approx 4 \times 10^4 \). Fig. \ref{fig7} illustrates the fidelities of 16 computational basis states resulting from the application of the second protocol for two photons in two DoFs, against 16D input states under the parameters \( g/\kappa = 2.4 \), \( \kappa_s/\kappa = 0 \), and \( \gamma = 0.1\kappa \). As evidenced, the fidelity for each computational basis exceeds 0.9999.

\section{Summary}\label{sec6}

In summary, we propose three practical protocols for realizing the 16D CSUM gate. The first protocol solely relies on linear optical elements, with current optical technologies, yielding the efficiency $\eta_1$ of 1/9 and the fidelity $F_{1}$  greater than 0.9936 under realistic conditions delay error \( \Delta = \pi/36 \), imperfection of the transmission ratio \( \xi = 0.02 \), deviation in mirror mounts $\phi=0.001$, and polarization extinction ratio  $p=0.0001$. The second protocol utilizes the photon scattering characteristics of the microcavity-QD system with the efficiency $\eta_2$ of  0.9913 and the fidelity $F_{2}$ of  0.9999  keeping the condition side leakage rate $\kappa_s/\kappa=0$ and coupling
strength $g/\kappa = 2.4$. As the error-heralded mechanism  is incorporated for the third protocol, under the same
condition as the second protocol, its efficiency $\eta_3$ of 0.9870 and  fidelity $F_{3}$ of 1. Moreover,
this encoding strategy of the 16D CSUM gates reduce resource overhead, suppress decoherence, and shorten operational timescales, thereby facilitating efficient storage and enabling ultrafast QIP. The proposed protocols  provide a scalable framework for advancing HD quantum computing, delivering enhanced processing capabilities and robust performance.

	\medskip
		
	\medskip
	\textbf{Acknowledgements} \par 
	This work was supported in part by Natural Science Foundation of China under Contract 61901420; in part by Fundamental Research Program of Shanxi Province under Contract 20230302121116.
	
	\medskip
	\textbf{Conflict of Interest} \par
	The authors declare no conflict of interest.
	
	\medskip
	\textbf{Data Availability Statement} \par
	The data that support the findings of this study are available from the corresponding author upon reasonable request.
	
	\bibliographystyle{MSP}
	\bibliography{document}
	\medskip
\end{document}